\documentclass[article,prc,notitlepage]{revtex4-1}

\usepackage{amsfonts,amssymb,amsmath}  
\usepackage{array,dcolumn}             
\usepackage{longtable}                 
\usepackage{xcolor}

\usepackage{commath}                   
\usepackage{xspace}                    
\usepackage{graphicx}                  

\begin{document}

\title{Implications of Missing Resonances in Heavy Ions Collisions}
\date{\today}

\author{Jacquelyn Noronha-Hostler}
\affiliation{Department of Physics, University of Houston, Houston TX 77204, USA}

\begin{abstract}
Between 2004 until today many new resonances in the QCD spectrum have been added by the Particle Data Group.  However, it is still possible that there are further states that have yet to be measured, as predicted by Lattice QCD calculations and Quark Models. In this paper I review how these missing states would influence the field of heavy-ion collisions.  Additionally, the quantum numbers that characterize these missing states are also discussed.
\end{abstract}
\maketitle

\section{Introduction}

Efforts at Jefferson Laboratory have resulted in a significant increase in the number of known QCD resonances since 2004, which are now listed in the Particle Data Group (PDG). Furthermore, various theory groups have predicted even further states using Lattice QCD \cite{Dudek:2009qf} and quark models \cite{Capstick:1986bm,Ebert:2009ub}, giving marginal support to an exponentially increasing mass spectrum \cite{Hagedorn:1965st} up to a given value of mass.  Even amongst the resonances listed within the PDG some are more confidentially measured than others.  Thus, a star system rating was developed where states with the highest confidence level are listed as **** states and those with the least are * states.

It is natural to wonder how many more states are still missing and what is the top possible maximum mass of these resonances.  The implications of these missing resonances have been studied extensively over the last 15 years in the field of heavy-ion collisions. Results that have garnered the most attention include the effect of missing resonances on the transport coefficients in the hadron gas phase \cite{NoronhaHostler:2008ju} as well as the possibility that there are missing strange baryons found using Lattice QCD \cite{Bazavov:2014xya}.  This proceedings reviews some of the main findings in the field of heavy-ion collisions and also discusses the mass range and the quantum numbers within which further missing states may be measured.

\section{Comparisons with Lattice QCD}

At the transition region between the hadron gas phase into the Quark Gluon Plasma, Lattice Quantum Chromodynamics (QCD) has been enormously successful in calculating thermodynamic quantities. In the early 2000's the hadron resonance gas model (where it is assumed that an interacting gas of hadrons can be described by a gas of non-interacting hadrons and their resonances) was used to calculate thermodynamic quantities. For instance, the pressure of the hadron resonance gas model is
\begin{equation}
p/T^4= \sum_{i}^{PDG}\frac{d_i}{2\pi^2}\left(\frac{m_i}{T}\right)^2\sum_{k=1}^{\infty}\frac{(\pm1)^{k+1}}{k^2}K_2\left(\frac{km_i}{T}\right)	\cosh\left[\frac{k\left(B_i\mu_B+S_i\mu_S+Q_i\mu_Q\right)}{T}\right]
\end{equation}
where $d_i$ is the degeneracy, $m_i$ is the mass,  $B_i$ is the baryon number, $S_i$ is the strangeness, and $Q_i$ is the electric charge of each individual hadron taken into account.  Thus, the only real variable in this model is the total number of resonances as dictated by the PDG.

Due to having fewer experimentally measured resonances in the early 2000's, the hadron gas model was unable to match Lattice QCD calculations of the pressure, energy density and entropy \cite{Karsch:2003vd} (an unphysical pion mass was also an issue on the lattice side but that has now been resolved).   Using the PDG lists from 2004 predictions were made that missing states would improve the fits to Lattice QCD results \cite{NoronhaHostler:2008ju,Majumder:2010ik,NoronhaHostler:2012ug}, which were later confirmed \cite{Borsanyi:2013bia} when the PDG's 2014 list \cite {Agashe:2014kda} was released. However, more massive resonances should not contribute strongly to the total pressure and, thus, if further missing states are measured it is very unlikely that they would affect the comparisons with Lattice QCD thermodynamic quantities.  Additionally, such comparisons make it very difficult to determine the quantum numbers of missing resonances and, thus, more differential observables from Lattice QCD are currently being introduced. As a final note, it has also been postulated that missing states can assist in understanding phase changes from hadronic to deconfined matter and the order of the phase transition \cite{Moretto:2005iz,Begun:2009an,Zakout:2006zj,Ferroni:2008ej,Bugaev:2008iu,Ivanytskyi:2012yx}.

\section{Searching for missing resonances assuming an exponential mass spectrum}\label{sec:HAG}

In the 1960's Hagedorn \cite{Hagedorn:1965st} suggested that QCD resonances followed an exponentially increasing mass spectrum.   In his picture, there was a limiting temperature $T_H$ above which as one adds in more  energy to the system the temperature would not increase but rather it would open up new degrees of freedom. 
\begin{figure} [h]
  \centering
  \includegraphics[width=0.4\textwidth]{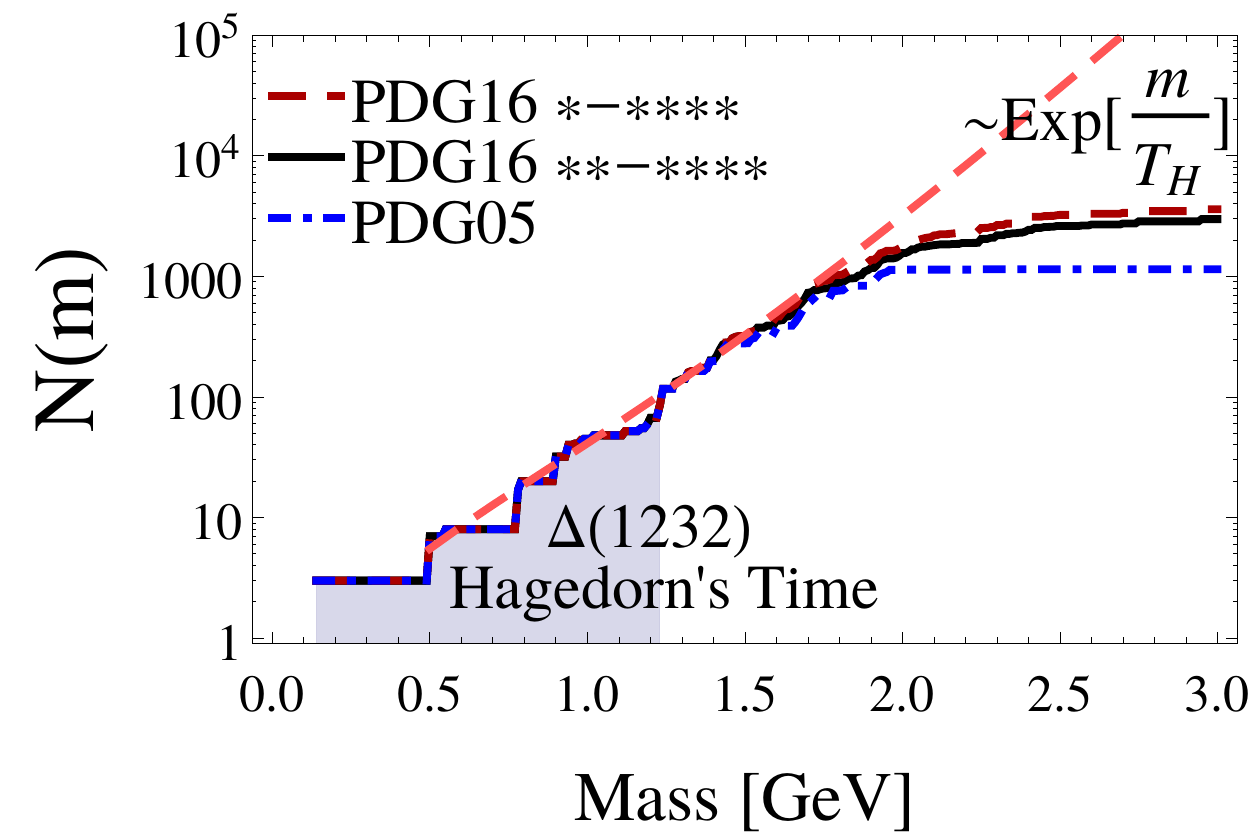}
  \caption{(Color online) Mass spectrum of all known (light and strange) hadrons as listed in the PDG from 2004 \cite{Eidelman:2004wy} and 2016 \cite{Olive:2016xmw} either for more well-measured states (**-****) or all possibly measured states (*-****). }
  \label{fig:all}
\end{figure}
In Hagedorn's time resonances were only measured up to $\Delta(1232)$. However, in the most recent release from the Particle Data Group (PDG) \cite{Olive:2016xmw} (light and strange) hadrons were measured up to $m\sim 2.5$ GeV.  One can see in Fig.\ \ref{fig:all} that the mass range that demonstrates an exponentially increasing mass spectrum has expanded significantly from the 1960's and, in fact, even between 2004 to 2016 many new states have been measured (although reservations regarding an exponential mass spectrum were also discussed here \cite{Cohen:2011cr}).

Returning to  Fig.\ \ref{fig:all}, the mass spectrum of all light and strange hadrons are plotted versus their mass such that:
\begin{equation}
N(m)=\sum_i d_i \Theta(m-m_i)
\end{equation} 
where $d_i$ is their degeneracy and $m_i$ is the mass of the $i^{th}$ resonance.  The exponential fit made in Fig.\ \ref{fig:all} is based upon Hagedorn's Ansatz for the degeneracy of ``Hagedorn States" (i.e. the massive states missing from the exponentially increase mass spectrum):
\begin{equation}
\rho(M)=\int_{M_{0}}^{M}\frac{A}{\left[m^2 +(m_{0})^2\right]^{a}}e^{\frac{m}{T_{H}}}dm
\end{equation}
where $m_0$ is the mass of the lowest stable particle of the spectrum, $T_{H}=165$ MeV is the Hagedorn temperature such that $T_c \leq T_H$ (a discussion on the choice of $T_H$ follows below), $M_0$ is the minimum mass where Hagedorn states begin, and $A$ is a free parameter. The power of $a$ has a non-trivial affect such that when $a=3/2$ Hagedorn states are more likely to decay into two body decays and when $a=5/4$ multi-body decays are more likely \cite{Frautschi:1971ij}.  I assume that multi-body decays are more likely with heavier resonances (and a brief glance through the PDG booklet confirms this assumption) so $a=5/4$ is taken.  

\begin{figure} [h]
\begin{tabular}{c c}
  \includegraphics[width=0.4\textwidth]{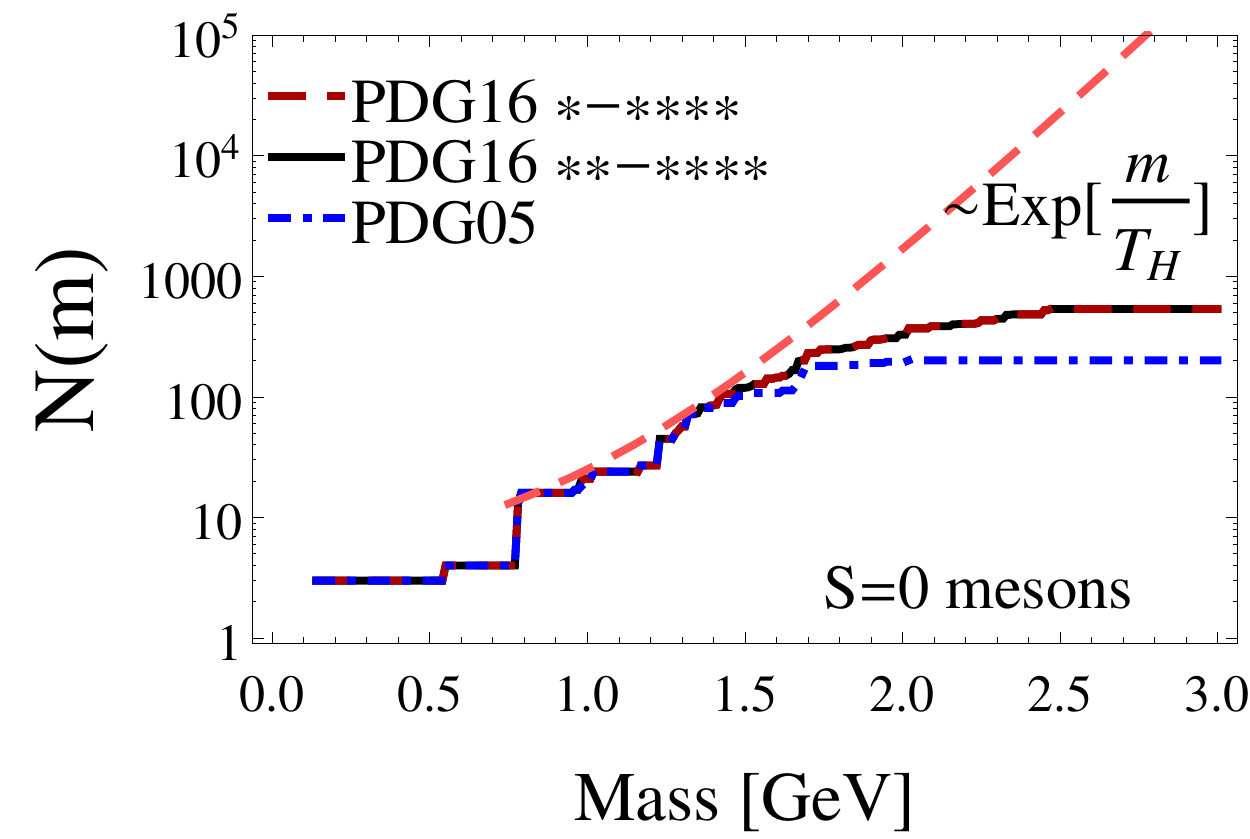} &  \includegraphics[width=0.4\textwidth]{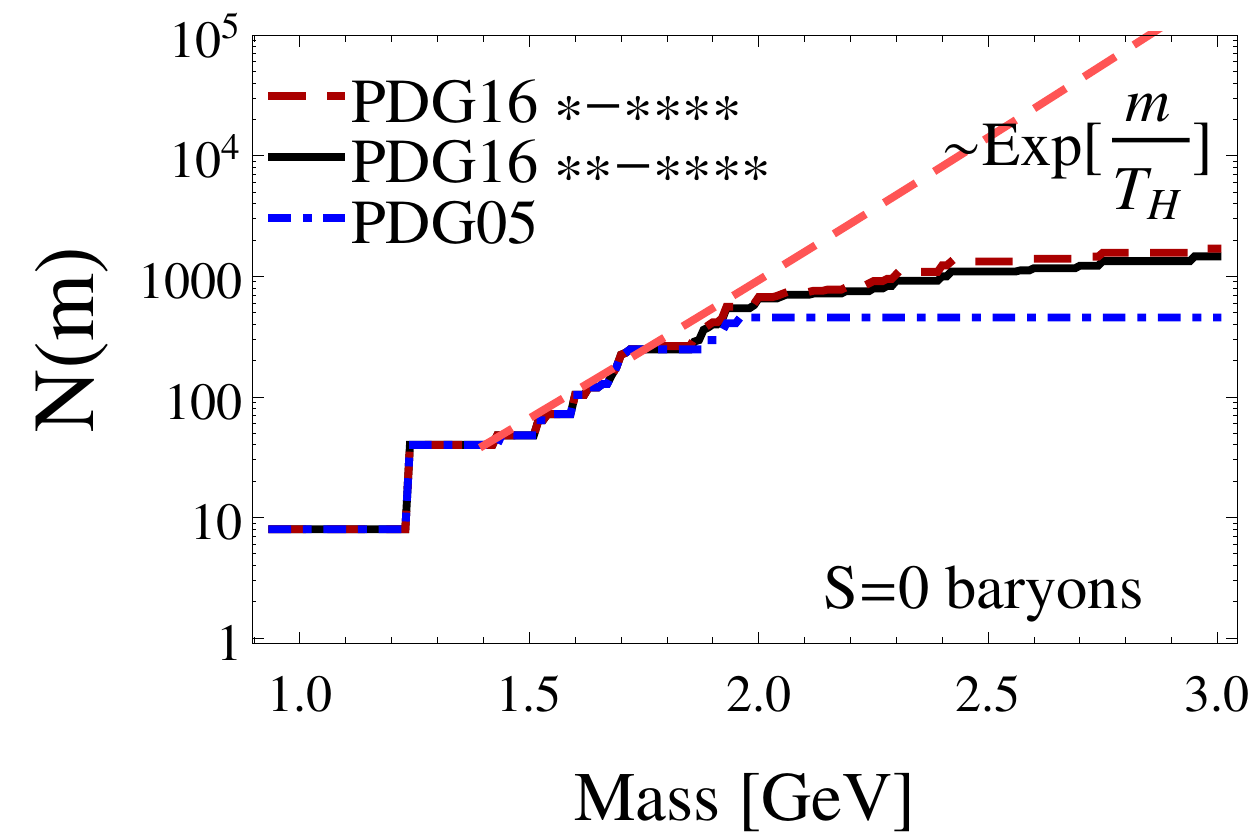} 
\end{tabular}
  \caption{(Color online) Mass spectrum of all known light mesons (left) and light baryons (right) as listed in the PDG from 2004 \cite{Eidelman:2004wy} and 2016 \cite{Olive:2016xmw} either for more well-measured states (**-****) or all possibly measured states (*-****). }
  \label{fig:light}
\end{figure}

In \cite{Broniowski:2004yh,Chatterjee:2009km,Lo:2015cca} the known hadrons were separated by families (light mesons, strange mesons etc) in order to extract their individual mass spectra.  Here I return to this approach to determine:
\begin{itemize}
\item if all families see an exponentially increasing mass spectrum,
\item if there are any gaps in the mass spectrum that could possibly be filled with missing states,
\item if there are variations in the needed Hagedorn temperature by species (following the idea in \cite{Bellwied:2013cta,Noronha-Hostler:2016rpd}).
\end{itemize} 
In Fig.\ \ref{fig:light} the mass spectra of the light mesons and baryons are shown where both are fitted to a temperature of $T_H=155$ MeV, which is almost exactly what one would expect from Lattice QCD.  Light mesons begin to deviate from an exponential mass spectrum at around $M\sim 1.5$ GeV whereas light baryons deviate around $M\sim 2$ GeV. In both cases a visible improvement was seen going from the PDG 2004 to 2016.  For light hadrons the difference between *-**** states and **-**** states are negligible.  

\begin{figure} [h]
\begin{tabular}{c c}
  \includegraphics[width=0.4\textwidth]{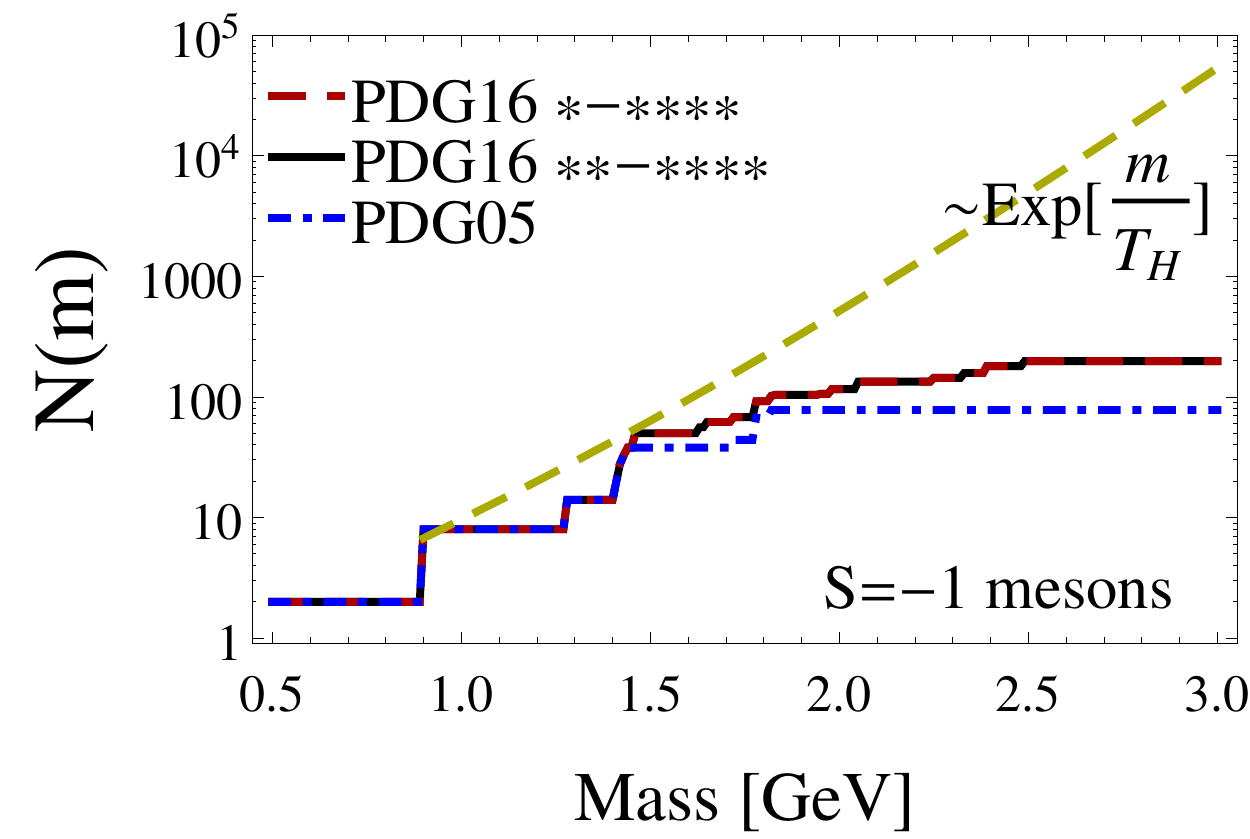} &  \includegraphics[width=0.4\textwidth]{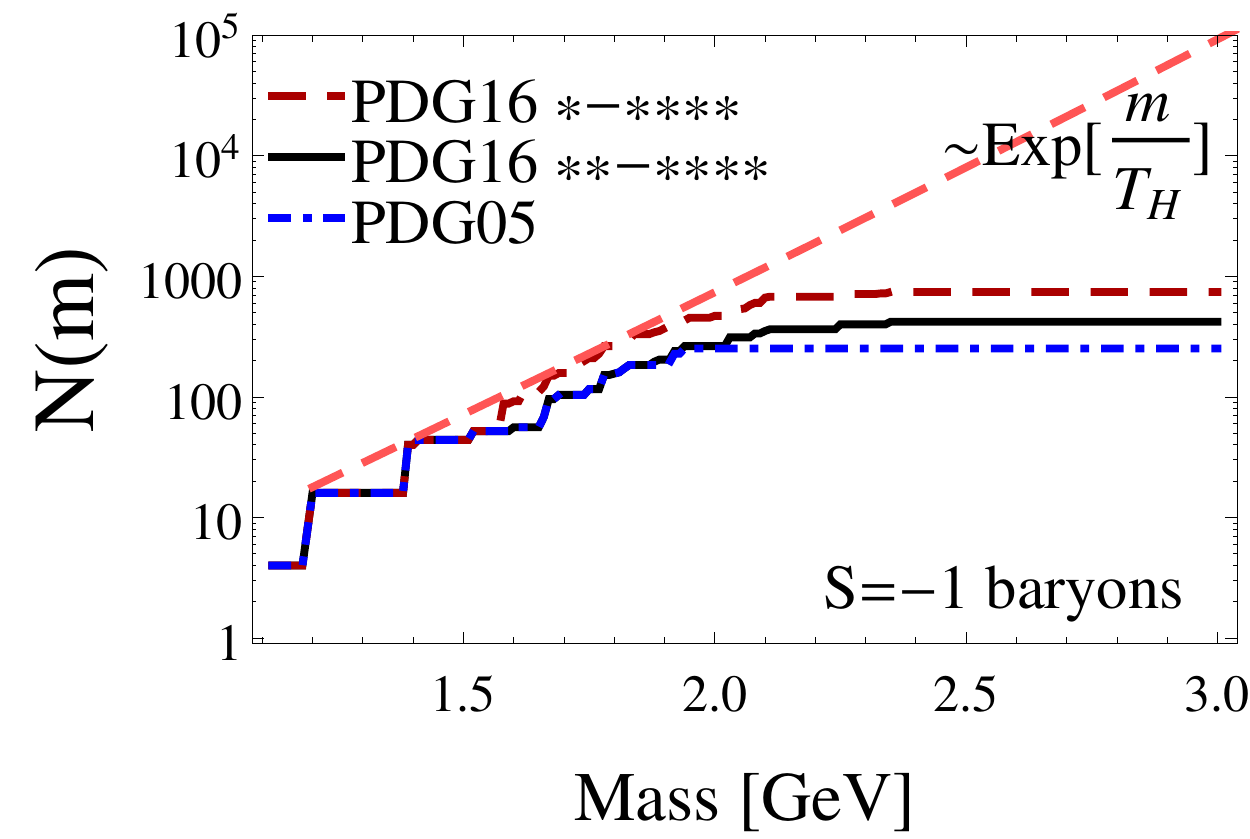} \\
    \includegraphics[width=0.4\textwidth]{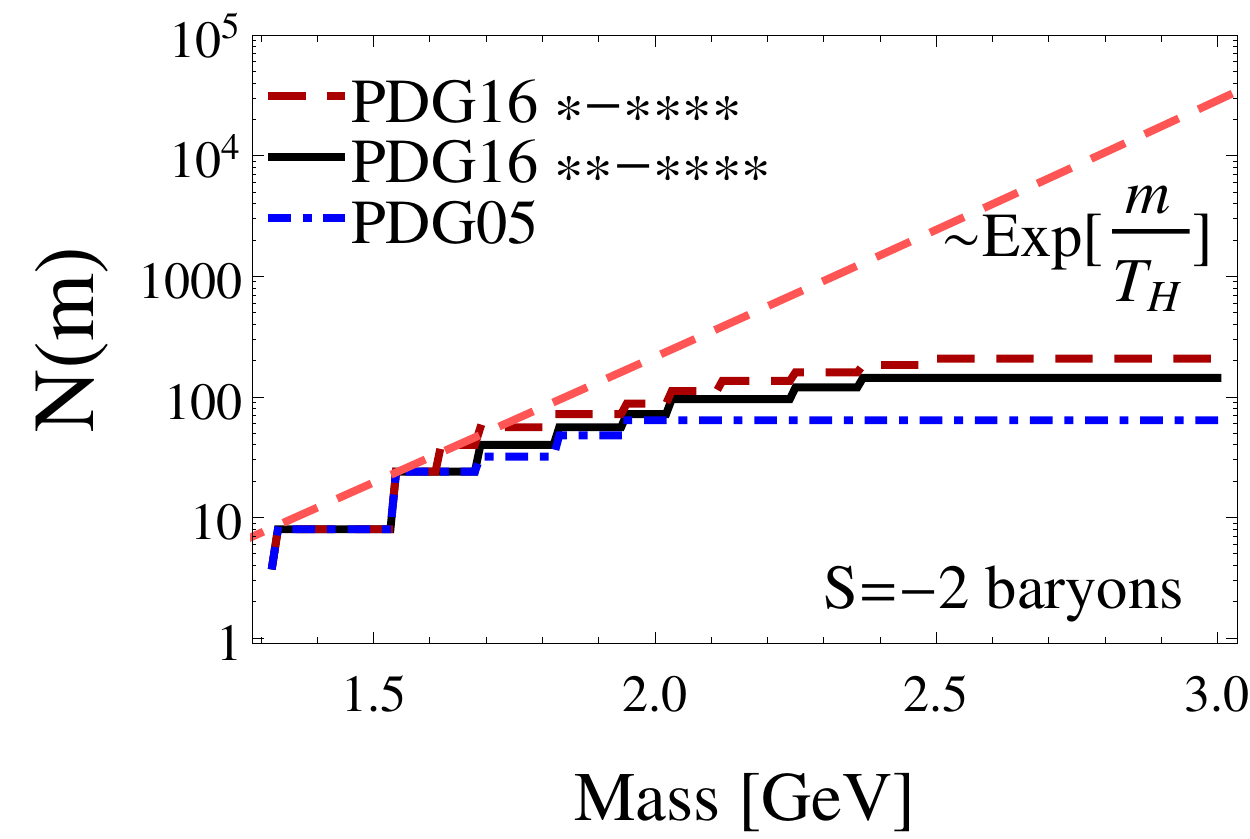} &  \includegraphics[width=0.4\textwidth]{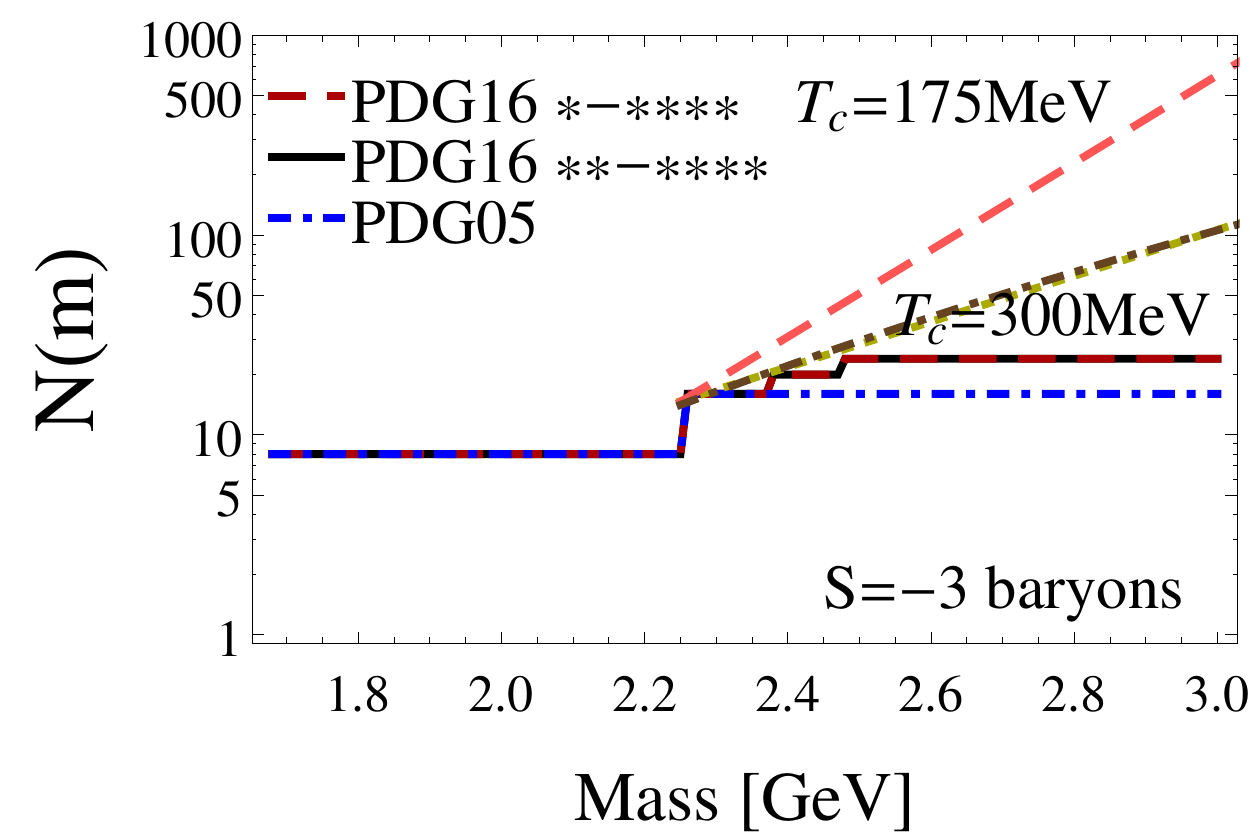} \\
\end{tabular}
  \caption{(Color online) Mass spectrum of all known strange hadrons (mesons and baryons $|S|=1-3$) as listed in the PDG from 2004 \cite{Eidelman:2004wy} and 2016 \cite{Olive:2016xmw} either for more well-measured states (**-****) or all possibly measured states (*-****). }
  \label{fig:strange}
\end{figure}

In Fig.\ \ref{fig:strange} the strange mesons and baryons are shown by their strangeness content.  Unlike for the light hadrons, a higher Hagedorn temperature of $T_H=175$ MeV is needed in order to fit the mass spectra (the one exception being Omega baryons), which is in-line with the possibility that strange particles have a higher transition temperature than light ones.  While $|S|=1$ and   $|S|=2$ baryons work well with an exponential mass spectrum up to $M\sim 2$ GeV and $M\sim 1.7$ GeV, respectively, the strange mesons and Omega baryons do not fit well to exponential mass spectra.  In the case of strange mesons, it appears that there is a large gap in the $M=1-1.5$ GeV region.  For Omega baryons there are so few states that the only possibility is to fit the spectrum with a $T_H=300$ MeV because the $T_H=175$ MeV was significantly too stiff, which is well above the expected values.  Thus, looking for mid-mass strange mesons and $|S|=3$ baryons are likely candidates for missing states.  Additionally, $|S|=1$ and   $|S|=2$ may have missing states but most likely only in the more massive region $M\gtrsim 2$ GeV. 

\section{Freeze-out Parameters and Chemical Equilibration Time}

In heavy-ion collisions, the system size is extremely small and is constantly expanding (at almost the speed of light) and cooling down.  Thus, due to dynamical effects, hadrons may not manage to reach chemical equilibrium on such short time scales.  However, due to the success of thermal fits \cite{Andronic:2005yp}, many believe that even on such short time scales chemical equilibrium can be reached.  Studies found that $2\leftrightarrow 2$ body reactions were not able to reach chemical equilibrium on such short time scales, especially at SPS energies and higher.  It was then suggested that multi-mesonic reactions could speed up chemical equilibration times \cite{Rapp:2000gy,Kapusta:2002pg} to explain SPS energies, however, with the known resonances it was still not enough to explain RHIC data  \cite{Heinz:2006ur}. To describe RHIC energies extra, massive resonances were needed that had large decay widths and could act as a catalyst to speed up hadronic reactions \cite{Greiner:2004vm,Pal:2005rb,NoronhaHostler:2007fg,NoronhaHostler:2007jf,NoronhaHostler:2009cf,Pal:2013oha,Beitel:2014kza,Beitel:2016ghw}.

\begin{figure} [h]
\begin{tabular}{c c}
  \includegraphics[width=0.4\textwidth]{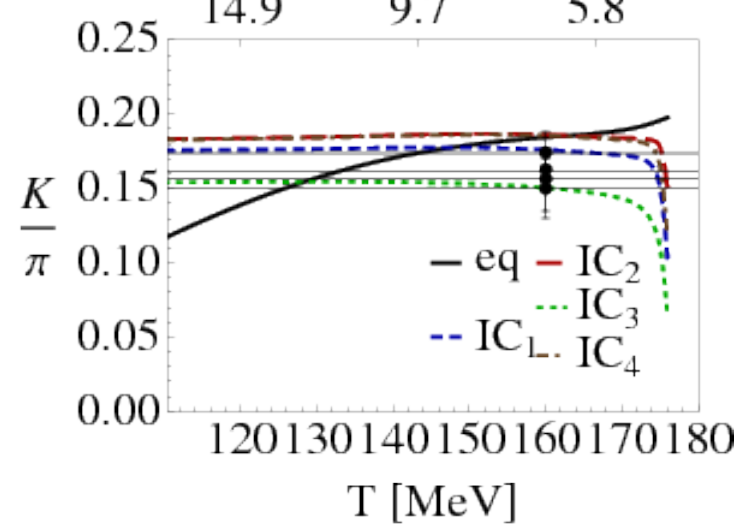} &  \includegraphics[width=0.5\textwidth]{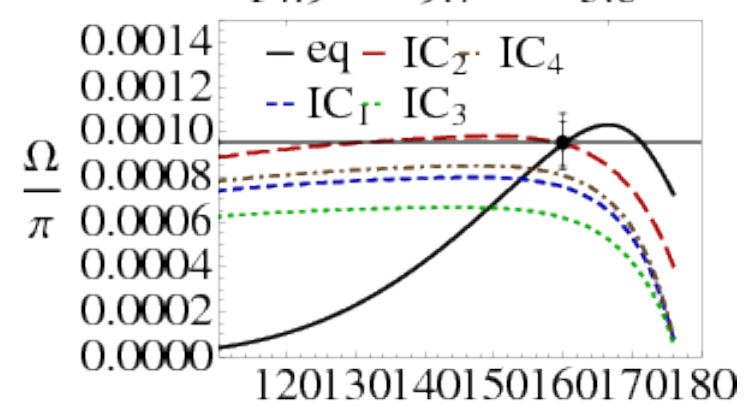} 
\end{tabular}
  \caption{(Color online) $K/\pi$ ratio and $\Omega/\pi$ ratio in a cooling, expanding fireball where the reactions are driven by Hagedorn states \cite{NoronhaHostler:2007jf,NoronhaHostler:2009cf} compared to RHIC data. }
  \label{fig:dyn}
\end{figure}
\begin{figure} [h]
  \centering
  \includegraphics[width=0.5\textwidth]{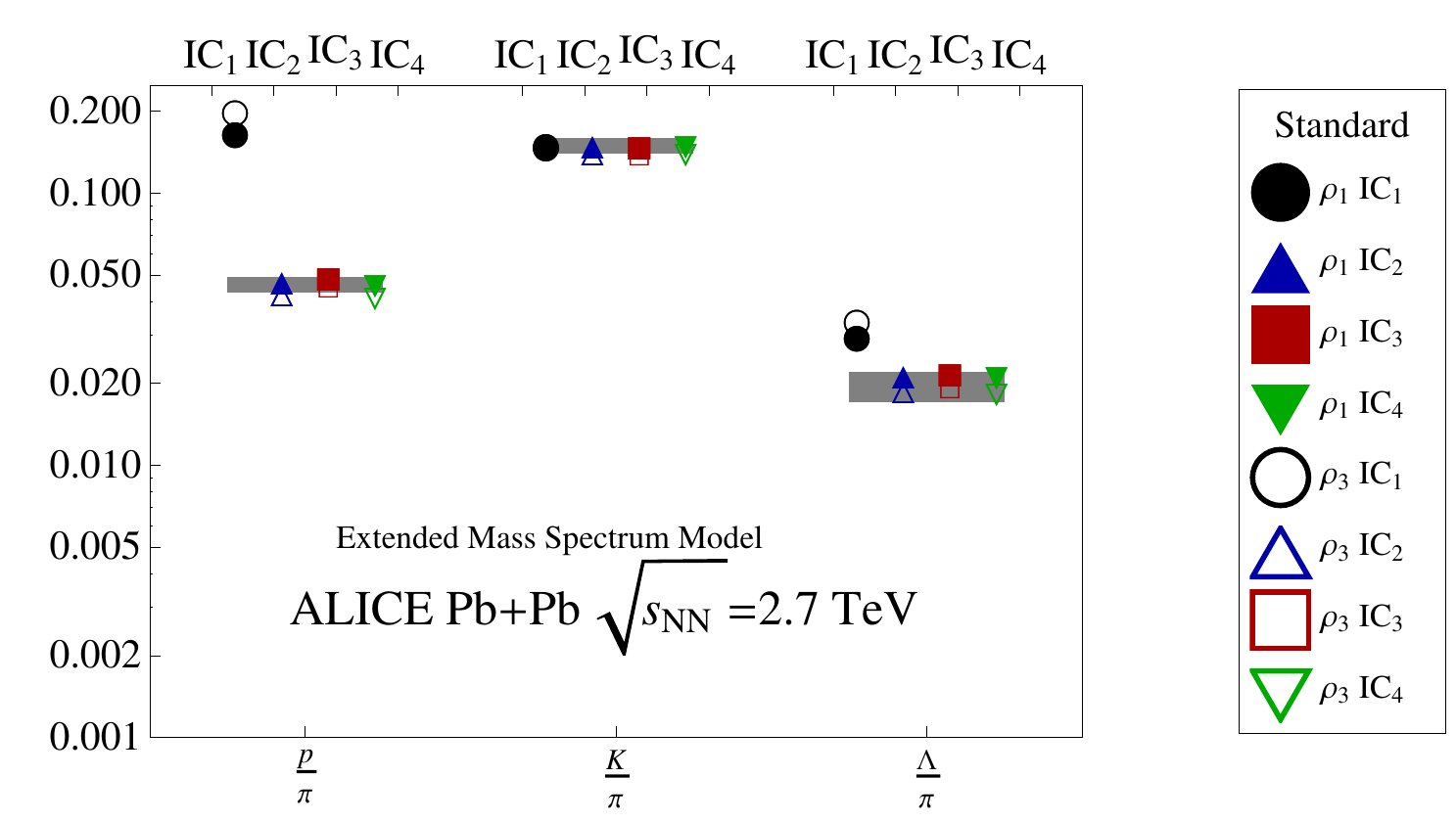}
  \caption{(Color online) Particle ratios calculated within a dynamical multi-body interaction model using Hagedorn states \cite{NoronhaHostler:2007jf,NoronhaHostler:2009cf} compared to ALICE data.}
  \label{fig:sum}
\end{figure}

In Fig. \ref{fig:dyn} the description of these missing states from \cite{NoronhaHostler:2007jf,NoronhaHostler:2009cf} is shown within a cooling, expanding fireball compared to experimental data from RHIC.   Because all reactions are dynamical, there is not a set ``freeze-out temperature" but each species is allowed to reach chemical equilibrium on its own, which varies according to the description of the Hagedorn Spectrum, branching ratios, and initial conditions. A summary of the final particle ratios is shown in Fig.\ \ref{fig:sum} compared to LHC experimental particle ratios. 

Further studies have investigated if these missing resonances could affect the freeze-out temperature of hadrons \cite{NoronhaHostler:2009tz,Bazavov:2014xya} i.e. at what temperature hadrons reach chemical equilibrium. However, it is unlikely that additional states beyond the PDG 2004/2016 will strongly affect the freeze-out temperature \cite{Noronha-Hostler:2016sje}.  Additionally, they were proposed as a solution for the tension between strange and light particles yields in theoretical models \cite{Bazavov:2014xya,Noronha-Hostler:2014usa,Noronha-Hostler:2014aia} though another likely solution may be a difference between the light and strange freeze-out temperatures \cite{Bellwied:2013cta,Noronha-Hostler:2016rpd}. 

\section{Transport Coefficients and Hydrodynamics}

\begin{figure} [h]
\begin{tabular}{c c}
  \includegraphics[width=0.4\textwidth]{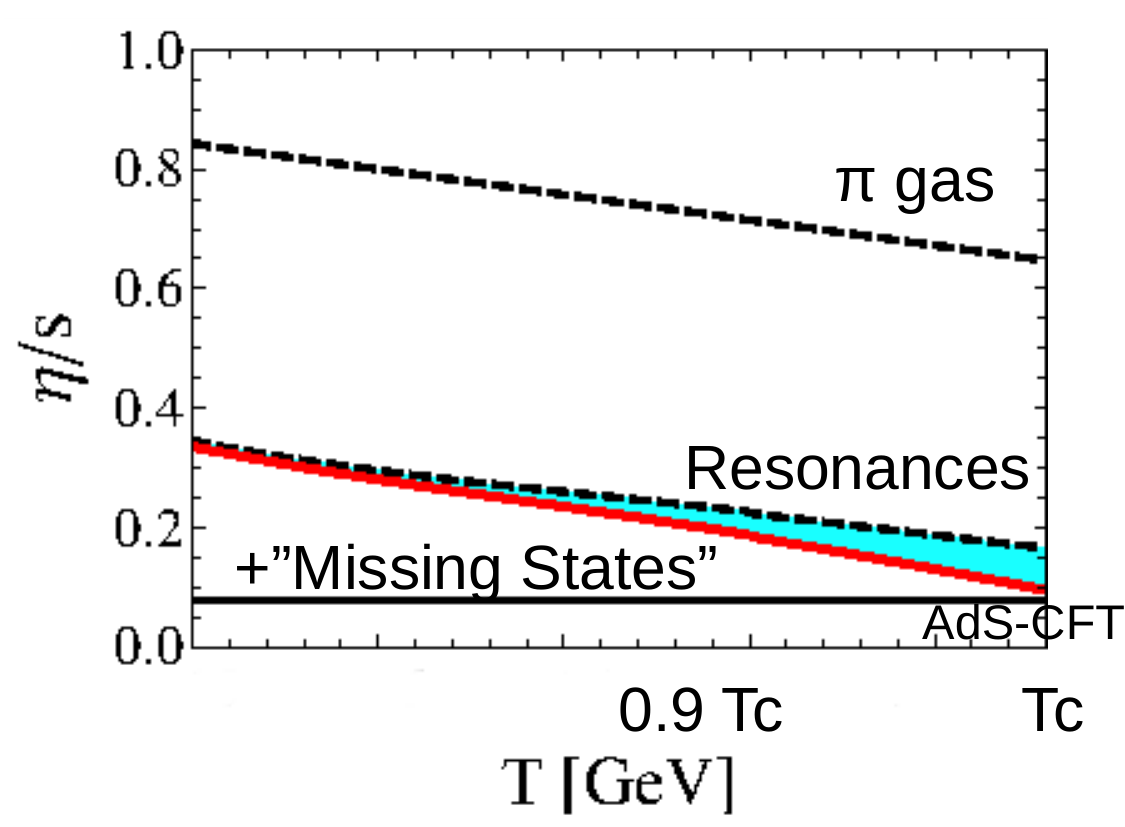} &  \includegraphics[width=0.4\textwidth]{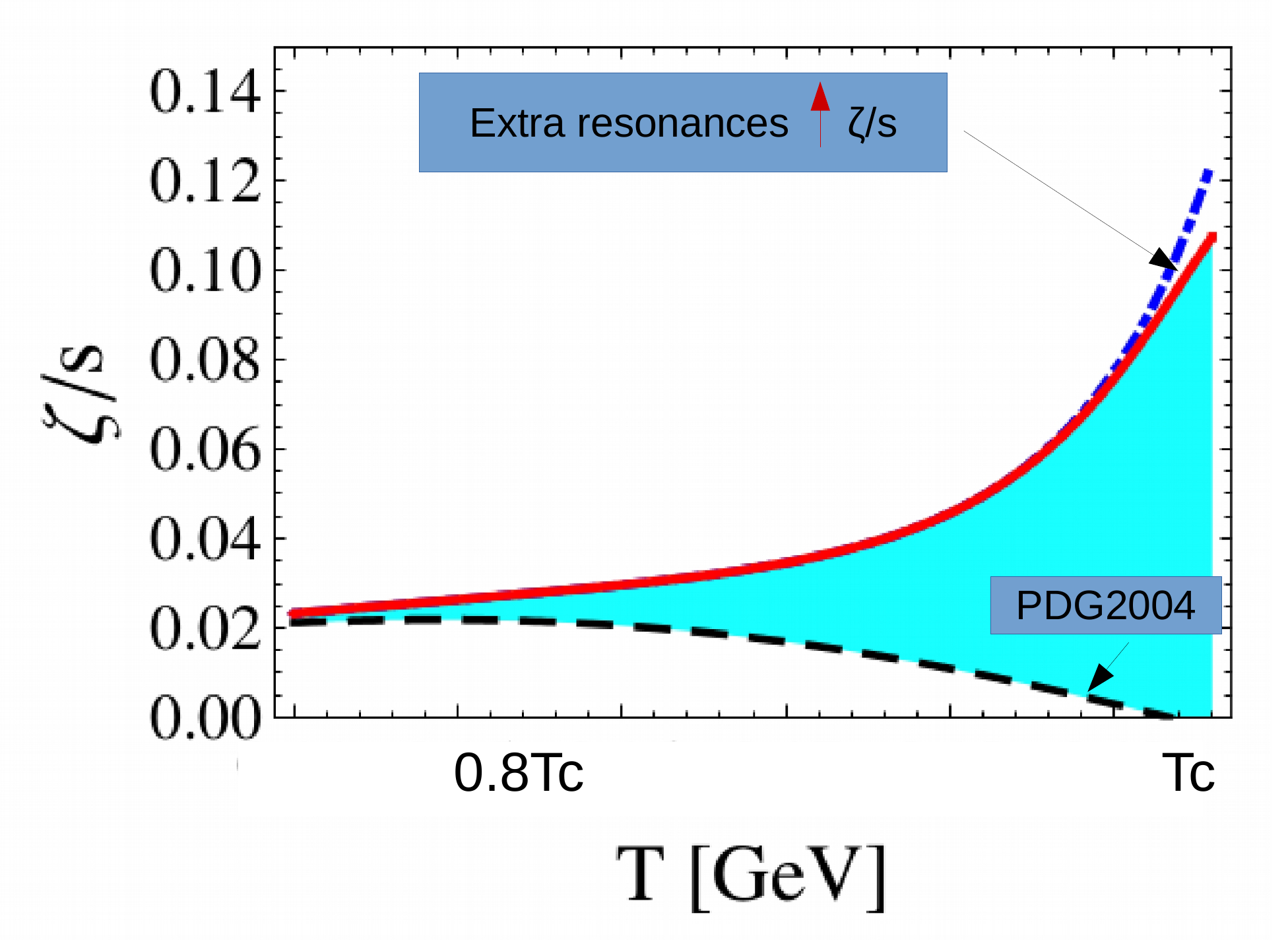} 
\end{tabular}
  \caption{(Color online) The effect of adding in missing resonances to the shear viscosity to entropy density ratio (left) and bulk viscosity to entropy density ratio (right) within the hadron gas phase  \cite{NoronhaHostler:2008ju}.  }
  \label{fig:trans}
\end{figure}

One of the most significant findings in heavy-ion collisions is that the Quark Gluon Plasma is a nearly perfect fluid (extremely small viscous effects), which can be well described by relativistic viscous hydrodynamics (for a review, see \cite{Heinz:2013th}).  Adding in missing resonances helped to explain the extremely low shear viscosity to entropy density ratio expected at the phase transition between the hadron gas phase and the Quark Gluon Plasma \cite{NoronhaHostler:2008ju,NoronhaHostler:2012ug} as shown  on the left in Fig.\ \ref{fig:trans}. This calculation also gave support to the presence of a peak in the bulk viscosity \cite{Karsch:2007jc} at the phase transition region, as shown on the right in Fig.\ \ref{fig:trans}.

The major signature of perfect fluidity in heavy-ion collisions is known as elliptical flow, $v_2$.  Elliptical flow is the second Fourier coefficient of the particle spectra and indicates that there is a strong (spatial) elliptical shape in the initial conditions, which is turned into an elliptical shape in momentum space due to the Quark Gluon Plasma's nearly perfect fluid-like nature.  If the Quark Gluon Plasma did not act as a fluid or if it was very viscous than $v_2\sim 0$.

\begin{figure} [h]
\begin{tabular}{c c}
  \includegraphics[width=0.35\textwidth]{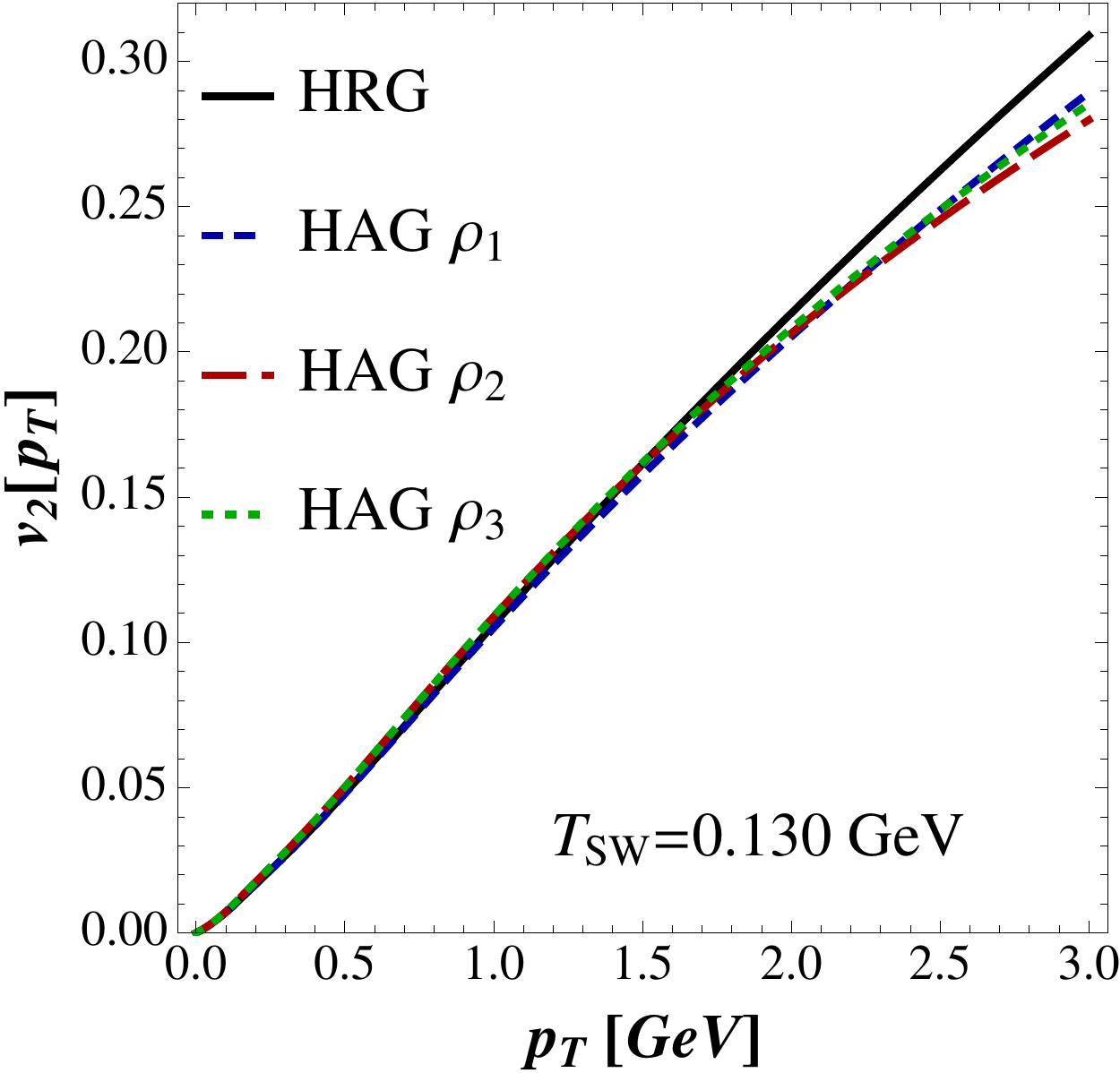} &  \includegraphics[width=0.35\textwidth]{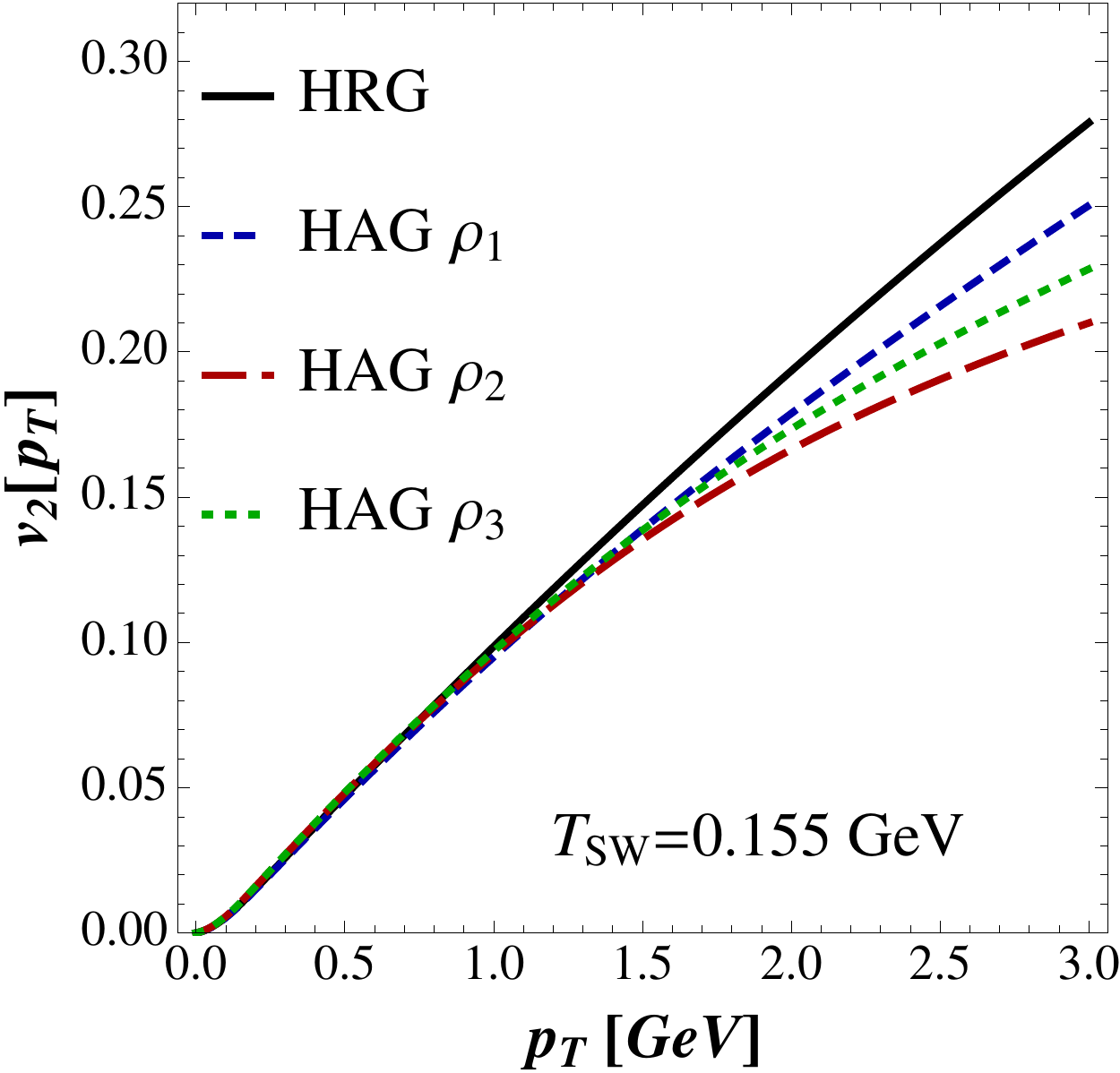} 
\end{tabular}
  \caption{(Color online) Effect of different descriptions of Hagedorn states on the elliptical flow at $T_{SW}=130$ MeV (left) and $T_{SW}=155$ MeV (right) from \cite{Noronha-Hostler:2013rcw}. Solid line includes all PDG2014 resonances.}
  \label{fig:v2}
\end{figure}
In \cite{Noronha-Hostler:2013rcw} the effects of adding in missing states were studied on $v_2$. For high momenta and higher freeze-out temperatures extra resonances suppressed elliptical flow. Thus, missing states not included in hydrodynamical models introduce a systematic error in comparisons of $v_2$ to experimental data.  Recently, the hadron gas phase was shown to play a significant role in photon production \cite{Paquet:2015lta} though theoretical calculations still underpredict experimental photon yields. It would be interesting to see if effects from missing resonances play a role in solving the photon ``puzzle".

\section{Conclusions}

In conclusion, missing resonances can affect various aspects of heavy-ion collisions.  The dynamics of the hadron gas phase plays a crucial role in comparisons between theoretical models and experimental data. If there are missing resonances, they contribute as a source of systematic error in all theoretical calculations.  Using an exponential mass spectra, there may be a gap in the strange mesons between $m=1-1.5$ GeV, which is a potential region to search for new resonances.   Therefore, strange hadron observables may be especially sensitive to missing states. 

Due to the sensitivity of heavy-ion collisions to missing resonances, a strong collaboration between the fields of hadron spectroscopy and heavy-ion collisions could be extremely fruitful.  For instance, the creation of a database that included all the known PDG resonances (and eventually also states predicted by Lattice QCD or Quark Models) with all their characteristic information (degeneracy, mass, quantum numbers, decay channels, etc) sorted by their star rating in a format that is compatible with heavy-ion codes, would significantly speed up comparisons between the fields and allow for systematic checks on which resonances influence heavy-ion observables the most. 

\section{Acknowledgements}

This  work is supported by the National Science Foundation under grant no. PHY-1513864. 

\bibliography{library}
\end{document}